\documentclass[pra,showpacs]{revtex4}
\usepackage{amsmath,amssymb,latexsym}
\usepackage{newlfont}
\usepackage{amsfonts}

\newcommand{\bra}[1]{\ensuremath{\langle #1 |}}
\newcommand{\ket}[1]{\ensuremath{|\, #1 \rangle}}
\newcommand{\bk}[2]{\ensuremath{\langle #1 | #2 \rangle}}
\newcommand{\kb}[2]{\ensuremath{| #1 \rangle\!\langle #2 |}}

\newtheorem{definition}{Definition}
\newtheorem{theorem}{Theorem}
\newtheorem{corrolary}{Corrolary}

\begin{document}
\title{Separability, entanglement and full families of commuting normal
matrices}
\author{Jan Samsonowicz$^{1}$, Marek Ku\'s$^{2}$,
and Maciej Lewenstein$^{3}$}
\affiliation{$^1$Warsaw University of Technology, Faculty of Mathematics
and Information Science, Pl.\ Politechniki 1, 00-61 Warszawa, Poland\\
$^2$Center for Theoretical Physics PAS, Al. Lotnik\'ow 32/46,
02-668 Warszawa, Poland  \\
$^3$ICREA and ICFO - Institut de Ci\`{e}ncies Fot\`{o}niques,
Parc Mediterani de la Tecnologia,
Castelldefels, 08860 Spain}

\begin{abstract}
We reduce the question whether a given quantum mixed state is separable or
entangled to the problem of existence of a certain {\it full family of
commuting normal matrices} whose matrix elements are partially determined by
components of the pure states constituting a decomposition of the considered
mixture. The method reproduces many known entanglement and/or separability
criteria, and provides yet another geometrical characterization of mixed
separable states.

\end{abstract}

\email {J.Samsonowicz@alpha.mini.pw.edu.pl; Marek.Kus@cft.edu.pl; Maciej.Lewenstein@icfo.es}

\pacs{03.67.Mn,03.65.Ud}
\date{\today }
\maketitle


\section{Introduction}

\paragraph*{\bf Entanglement and separability problem.}
Entanglement is the most important quantum phenomenon, responsible for genuine,
distinct and unique properties of the quantum world, and possibilities this
world offers for future technological applications, such as  quantum
engineering, and quantum information \cite{books}. Despite enormous efforts,
many  fundamental questions concerning entanglement remain open (for an
excellent recent review see \cite{hor-rmp}, and \cite{bz06} for some general
geometric settings of the problem). In the seminal paper in 1989 Werner
\cite{werner} gave the definition of separable (i.e.\ non-entangled) states: a
state of a bi-partite system is separable iff it is a mixture of pure product
states. A simple question: given a state, is it separable or not, is known as
the {\it separability problem}. Only in very rare instances we know operational
sufficient and necessary criteria (SNC) that allow to solve this problem:
\begin{itemize}
\item for $2\times 2$ (two qubit) and $2\times 3$ (qubit--qutrit) systems
the SNC are given by the positive definiteness of the partial transform
\cite{hhh}; this is the famous PPT criterion, introduced by Peres as necessary
for separability in Ref.~\cite{Peres}.
\item for  3 qubit symmetric ("bosonic") states PPT criterion is also
SNC \cite{eckert}.
\item for continuous variables $1\times 1$ (one mode per party)  Gaussian
states,  PPT criterion (formulated at the level of correlation matrices) is a
SNC \cite{duan,simon}.
\item for continuous variables $m\times n$ (all bipartite) Gaussian states
there exist an operational SNC based on recursion for correlations matrices
\cite{giedke-all}.
\item for continuous variables tripartite $1\times 1\times 1$ Gaussian
states there exist an operational SNC based on "iteration" of PPT condition for
correlations matrices \cite{giedke-3}.
\end{itemize}
In general we have to rely either on only necessary criteria, or only
sufficient ones, or on numerical approaches. Although there exist very
efficient numerical procedures that employ optimization methods of
semi-definite programming \cite{doherty}, the complexity of the problem grows
with the dimensionality of the underlying Hilbert spaces: in fact it has been
proven that the problem belongs to the complexity NP-class \cite{gurvitz}.

\paragraph*{\bf Reformulations of the separability problem.}
The market for only necessary, or only sufficient criteria is growing
constantly, and it is impossible to review it in a non-review style  article
(for this reason we recommend the readers the review \cite{hor-rmp}). There are
also many attempts to reformulate the problem of separability in different
mathematical terms. A paradigm example for such an approach is the formulation
of the separability problem in terms of positive maps  due to Horodeccy
\cite{hhh}. A state is entangled iff there exist a positive map acting on, say,
Alice, such that when applied to the state in question, it produces a
non-positive definite operator. Similar approach deals with entanglement
witnesses, i.e. observables that have positive averages on all separable
states, but a negative average on some entangled state: A state is entangled
iff there exist a witness operator that detects it, i.e.\ has a negative
average. Obviously both these approaches are not operational, but nevertheless
they are extremely useful, since they allow to generate many  necessary
separability (sufficient entanglement) criteria via explicit construction of
positive maps \cite{maps}, witnesses \cite{witnesses}, and methods of their
(local) measurements (cf. \cite{detection}, respectively).

We have recently presented  another example of the reformulation of the
separability problem employing harmonic analysis on compact groups
\cite{korbicz}. In this approach quantum mechanical states are replaced by
non-commutative characteristic functions defined on the considered group, and
separability problem reduces to the question whether a characteristic function
defined on a product group of two groups can be represented as a mixture of
products of characteristic functions on each of the individual groups. The
present paper is in a sense similar to the Ref.~\cite{korbicz}: we present yet
another reformulation of the separability problem and reduce it to an
apparently unrelated question of existence of a {\it full} (in a sense
specified below) {\it family of commuting normal matrices}, whose matrix
elements are partially determined by components of the pure states that
constitute a decomposition of the considered mixed state.

\paragraph*{\bf Decompositions of mixed states.}
A given  genuine (not pure) mixed state $\rho$  has infinite  number of
decompositions in terms of projectors onto pure states. This fact has been
already recognized by Schr\"odinger in 1935 \cite{erwin}, and elaborated
thoroughly from the more modern view by Hughston {\it et al.} \cite{Hughston}.
Any decomposition of a density matrix of rank $r$ into $K$ projectors can be
described in terms of a rectangular $K\times r$ matrix, whose $r$ columns are
orthonormal. Such objects are known in geometry to form a so called
$V_{K,r}=U(K)/U(K-r)$ Stiefel manifold \cite{spivak}. Separability problem
might be also formulated as a problem of statistical mechanics of a fictitious
systems on the Stiefel manifold, characterized by a positive definite
Hamiltonian (cost-function) that vanishes for separable states \cite{jarek}.
Here, we follow another avenue: we consider all decomposition of $\rho$ into
$K$ terms for sufficiently large $K$: such decompositions are related via
unitary transformations $U(K)$. The matrix elements of the density matrix, on
the other hand, form a Gram matrix of scalar products of certain vectors from
this $K$ dimensional space. The first chapter of the paper is thus devoted to
the study of such Gram decompositions. It provides complementary results to the
Ref.~\cite{Hughston}.

\paragraph*{\bf Plan of the paper.}
As  stated above, Section II is devoted to the Gram decompositions, and its
main result is the Theorem 1, that describes how the two different Gram
decomposition are connected.  We present a reformulation of the separability
problem in Section III, in the Theorem 2. Here, an example of so called  Werner
matrices \cite{werner} is elaborated in detail. The section IV contains the
main result of this paper: a novel (but unfortunately not immediately
operational) SNC for separability in terms of existence of what we call a {\it
full family of commuting normal matrices (FFCNM)} (Theorem 3). We specify this
result to the particularly simple case of $2\times N$ systems, where the
separability SNC requires existence of a single normal matrix, whose matrix
elements are partially known. Here we use the general properties of the density
matrices in  $2\times N$ systems (as presented in Appendix) and formulate
elegant theorems on the existence of normal extensions  of partially known
matrices based on earlier and some new  results for $2\times 2$, and $2\times
3$. We discuss also application of our criteriou to PPT entangled states of
rank 5 in $2\times 4$ systems. We relate these results to the theory of
generalized concurrence \cite{kus}.

\section{Gram decompositions of density matrices}
\label{sec:gram}
\paragraph*{\bf Decompositions of density matrices.}
Physical states of composite quantum systems are represented by density
matrices, i.e.\ Hermitian, positive definite linear operators of trace one,
acting in the Hilbert space  $\mathcal H=\cal {H}_A \otimes \mathcal
{H}_B\otimes\ldots$, which is a tensor product of Hilbert spaces corresponding
to subsystems $A,B,\ldots$ of the considered system. In the following we shall
be concerned with states of bipartite systems in a finite-dimensional Hilbert
space i.e.\ described by positive definite Hermitian density matrices
$\rho=\rho^{\dagger}\ge 0$  with $\mathrm{Tr}\rho=1$, acting on the Hilbert
space of the composite system $\cal H=\mathcal {H}_A \otimes \mathcal {H}_B$.
Without loosing generality we will assume that dim ${\cal {H}}_A=M\ge 2$ and
dim ${\cal {H}}_B=N \ge M$, i.e.\ ${\mathcal
H}=\mathbb{C}^M\otimes\mathbb{C}^N=\mathbb{C}^{M\times N}$. In the following we
shall use the notation $r(\rho)$ for the rank of the matrix $\rho$ (which for
a Hermitian matrix equals the number of its nonvanishing eigenvalues).


Performing the spectral decomposition of $\rho$,
\begin{equation}\label{spectral}
\rho=\sum_{l=1}^r\lambda_{\,l}\kb{\psi_l}{\psi_l},
\end{equation}
where $\lambda_{\,l}$ are (positive) eigenvalues of $\rho$, $\ket{\psi_l}$ -
its eigenvectors and $r=r(\rho)$ - its rank. Defining
$\ket{\Psi_l}=\sqrt{\lambda_{\,l}}\ket{\psi_l}$ we thus decompose the
nonnegative-definite Hermitian matrix $\rho$ as a sum of rank-one operators,
\begin{equation}\label{rho}
\rho=\sum_{l=1}^r\kb{\Psi_l}{\Psi_l}.
\end{equation}
The decomposition of (\ref{rho}) into the sum a of rank-one operators is
non-unique. Indeed the vectors
\begin{equation}\label{woot}
\ket{\Phi_n}=\sum_{l=1}^rW_{nl}\ket{\Psi_l}, \quad n=1,\ldots,K\ge r,
\end{equation}
lead to another one,
\begin{equation}\label{rho1}
\rho=\sum_{n=1}^{K}\kb{\Phi_n}{\Phi_n},
\end{equation}
involving $K\ge r$ components, provided that the rectangular $K\times r$ matrix
$W$ fulfills $W^\dagger W=I$, where $I$ is the $r\times r$ identity matrix,
i.e. $W$ belongs to the $V_{K,r}$ manifold. In fact all possible decompositions
(\ref{rho1}) of $\rho$ into the sum of rank-one operators can be obtained from
the spectral decomposition of $\rho$ (\ref{rho}) in such a way
\cite{erwin,Hughston}.

\paragraph*{\bf Gram decompositions.}
Let $\{\ket{E_\nu}\}_{\nu=1,\ldots,M\times N}$ be a basis in
$\mathbb{C}^{M\times N}$. Starting form the spectral decomposition (\ref{rho})
we obtain for the matrix elements of $\rho$:
\begin{equation}\label{rhomunu}
\rho_{\mu\nu}=\bra{E_\mu}\rho\ket{E_\nu}=\sum_{l=1}^r\bk{E_\mu}{\Psi_l}\bk{\Psi_l}{E_\nu}
=\sum_{l=1}^r\overline{w}_\mu^{\,l}w_\nu^l=\langle w_\mu,w_\nu\rangle,
\end{equation}
\begin{equation}\label{w}
w_\nu:=
\left[
\begin{array}{c}
w_\nu^1 \\
w_\nu^2 \\
\vdots\\
w_\nu^r
\end{array}
\right]\in\mathbb{C}^r, \quad w_\nu^l:=\bk{\Psi_l}{E_\nu},
\end{equation}
where $\langle\cdot,\cdot\rangle$ is the standard Hermitian scalar product in
$\mathbb{C}^K$. It means that a Hermitian positive definite matrix is the Gram
matrix (i.e.\ the matrix of scalar products) of the vectors $w_\nu$ defined
above. A set of vectors $\{w_\nu\}$ fulfilling (\ref{rhomunu}) we will call a
Gram system for $\rho$, and we will say that it provides a Gram decomposition
of $\rho$. If we do not insist that the vectors $w_\nu$ are elements of
$\mathbb{C}^r$ we can construct other Gram decompositions of $\rho$, so in this
sense the Gram decomposition is non-unique. Indeed, defining
\begin{equation}\label{gram1}
w_{\nu}^\prime=Vw_{\nu},
\end{equation}
where $w_{\mu}^\prime\in\mathbb{C}^K$, with $K\ge r$ and
$V\in\mathbb{M}_{K\times r}$ - a rectangular matrix fulfilling $V^\dagger V=I$,
we have $\rho_{\mu\nu}=\langle w_\mu^\prime,w_\nu^\prime\rangle$. It is easy to
prove that all Gram decompositions of the matrix $\rho$ are obtained by the
transformation (\ref{gram1}) from the spectral one (\ref{rho}) and (\ref{w}).
In particular two Gram systems calculated from two decompositions (\ref{rho})
and (\ref{rho1}) are connected by the relation (\ref{gram1}) with
$V=\overline{W}$ (cf.\ (\ref{woot})). In the following we will also use the
fact that if two sets of vectors ${w_\nu^{\,\prime}}$ and
${w_\nu^{\,\prime\prime}}$, $\nu=1,\ldots,r$, $w_
\nu^{\,\prime}\in\mathbb{C}^K$, $w_\nu^{\,\prime\prime}\in\mathbb{C}^K$ are
Gram systems for the same positive-definite matrix $\rho$, i.e.\
$\left\langle w_\mu^{\,\prime},w_\nu^{\,\prime}\right\rangle=\left\langle
w_\mu^{\,\prime\prime},w_\nu^{\,\prime\prime}\right\rangle$, then there exists
a unitary $U$ acting in $\mathbb{C}^K$ such that
$w_\nu^{\,\prime\prime}=Uw_\nu^{\,\prime}$ for $\nu=1,\ldots,r$.

\paragraph*{\bf Gram decomposition in bipartite systems.}
Let us now take advantage of the fact that $\rho$ acts on a tensor product
space, i.e.\ we chose the basis $\{\ket{E_\nu}\}_{\nu=1,\ldots,M\times N}$ in
the form of product states
$\ket{E_\nu}=\ket{e_m}\otimes\ket{f_n}=:\ket{e_m\otimes f_n}$, $m=1,\ldots,M$,
$n=1,\ldots,N$ and repeat the calculation of (\ref{rhomunu}),
\begin{equation}\label{rhogen}
\rho_{ij,mn}=\bra{e_i\otimes f_j}\rho\ket{e_m\otimes f_n}
=\sum_{l=1}^r\left\langle e_i\otimes f_j\kb{\Psi_l}{\Psi_l} e_m\otimes
f_n\right\rangle
=\sum_{l=1}^r\overline{w}_{ij}^{\,l}w_{mn}^l=\langle w_{ij},w_{mn}\rangle,
\end{equation}
where now $w_{mn}$, $m=1,\ldots,M$, $n=1,\ldots,N$  are vectors in
$\mathbb{C}^r$, with components
\begin{equation}\label{wgen}
w_{mn}^l=\bk{\Psi_l}{e_m \otimes f_n}.
\end{equation}

If we assume that $\rho$ is of maximal rank $r=MN$ (which for $M\ge 3$ we take
for granted in the following), then $w_{mn}$ are linearly independent. In
particular, for any fixed $\widetilde{m}\in\{1,\ldots,M\}$
($\widetilde{n}\in\{1,\ldots,N\}$)  the vectors $w_{\widetilde{m}n}$,
$n=1,\ldots,N$ ($w_{m\widetilde{n}}$, $m=1,\ldots,M$) form a set of $N$ ($M$)
linearly independent vectors in $\mathbb{C}^K$, respectively. In the special
case $M=2$, the first of the latter statements can be also assumed to hold,
provided $\rho$ is (non-trivially) supported in the $2\times N$ space. If this
statement  was not true, the density matrix $\rho$ would have a product vector
in its kernel. In such situation either $\rho$ is entangled and not PPT, or if
it is PPT, than it can be represented as a mixture of a separable part and a
density matrix supported in $2\times (N-1)$ dimensional space (for proofs and
details see Ref.~\cite{2N}).

\paragraph*{\bf Relations between various Gram decompositions.}
Let $\{v_n\}_{n=1,\ldots,N}$ form an arbitrary set of linearly independent
vectors in $\mathbb{C}^K$ (in particular, in the light of the above remark, we
can choose $v_n=w_{1n}$). We can always find a family of linear maps
$F_m:\mathbb{C}^K\rightarrow\mathbb{C}^K$, $m=1,\ldots,M$, such that
\begin{equation}\label{Pm}
w_{mn}=F_mv_n,
\end{equation}
Note that such $F_m$'s
are uniquely defined only on the $N$-dimensional subspace of
 $\mathbb{C}^K$ spanned by $\{v_n\}_{n=1,\dots N}$.
If we apply the same procedure to the decomposition (\ref{rho1}) we will arrive at
\begin{equation}\label{Pm1}
w_{mn}^\prime = F_m^\prime v_n^\prime,
\end{equation}
with $v_n^\prime$, $n=1,\ldots,N$ some linearly independent vectors in
$\mathbb{C}^K$ and appropriate $F_m^\prime$.

The vectors $w_{mn}$ and $w_{mn}^\prime$ are connected by (\ref{gram1}), i.e.\
\begin{equation}\label{trans1}
F_m^\prime v_n^\prime=w_{mn}^\prime=Vw_{mn}=VF_mv_n.
\end{equation}
Since both sets $\{v_n\}$ and $\{v_n^\prime\}$, $n=1,\ldots,N$ are linearly
independent in, respectively, $\mathbb{C}^r$ and $\mathbb{C}^K$, there exists a
$K\times r$ matrix $\tilde{V}$ of maximal rank, such that $v_n^\prime=\tilde V
v_n$ for $n=1,\ldots,N$. Consequently
\begin{equation}\label{r1}
V^\dagger F_m^\prime\tilde{V} v_n=V^\dagger VF_m v_n=F_m v_n,
\end{equation}
where we used $V^\dagger V=I$.
We have thus shown the following
\begin{theorem}\label{theorem:gram1}
For the two decompositions of the Gram vectors of the form (\ref{Pm}) and
(\ref{Pm1}) stemming from two decompositions of $\rho$ into rank-one operators
of the form (\ref{rho}) and (\ref{rho1}) there exist two $K\times r$ matrices
$\tilde{V}$ nad $V$, the former of rank $r$ and the latter fulfilling
$V^\dagger V=I$, such that on the space spanned by (arbitrary chosen) $N$
linearly independent vectors $v_n\in\mathbb{C}^r$ the equality
\begin{equation}\label{trans2}
V^\dagger F_m^\prime\tilde{V}=F_m
\end{equation}
holds. \hfill $\Box$
\end{theorem}
Obviously $\tilde{V}$ depends on the choice of $\{v_n\}$.

\section{Separability problem}
\label{sec:separ}

\paragraph*{\bf Gram decompositions for separable states.}
Our goal is to characterize Gram decompositions for density matrices of
bipartite separable quantum systems. Recall that separable defined on $\cal
{H}_A\otimes\cal{H}_B$ systems are characterized by the following
\begin{definition} A state $\rho$ is separable if and only if
\begin{equation}\label{separable}
\rho=\sum_{i=1}^{k}p_i \rho_i^{A}\otimes \rho_i^B.
\end{equation}
where $\sum_ip_i=1$, $p_i\ge 0$, whereas $\rho_i^{A}$ and
$\rho_i^{B}$ are states on $\cal {H}_A$ and $\cal{H}_B$, respectively.
\end{definition} The above expression means that $\rho$ can be written as a
convex combination of product states.


To achieve the goal observe that performing a decomposition of the type
(\ref{rho}) for all matrices $\rho_i^{A}$ and $\rho_i^{B}$ in (\ref{separable})
and taking into account positivity of the coefficients $p_i$ we obtain that a
state $\rho$ is separable if and only if it can be decomposed in the form of
$K$ rank-one operators proportional to projections on simple tensors.
\begin{equation}\label{rhos}
\rho=\sum_{k=1}^K\kb{\varphi_k}{\varphi_k}\otimes\kb{\psi_k}{\psi_k}=
\sum_{k=1}^K\kb{\varphi_k\otimes\psi_k}{\varphi_k\otimes\psi_k},
\end{equation}
where $\ket{\varphi_k}\in\mathcal{H}_A$, $\ket{\psi_k}\in\mathcal{H}_B$.

Calculating matrix elements of $\rho$ in local bases
$\{\ket{e_i}\}_{i=1,\ldots,M}$ and $\{\ket{f_i}\}_{i=1,\ldots,N}$ in
$\mathcal{H}_A=\mathbb{C}^M$ and $\mathcal{H}_B=\mathbb{C}^N$, respectively, we
obtain
\begin{eqnarray}\label{rhoijmn}
\rho_{ij,mn}&=&\bra{e_i\otimes f_j}\rho\ket{e_m\otimes f_n}
\nonumber \\
&=&\sum_{l=1}^K\bk{e_i}{\varphi_l}\bk{f_j}{\psi_l}\bk{\varphi_l}{e_m}\bk{\psi_l}{f_n}
\nonumber \\
&=&\sum_{l=1}^K\overline{w}_{ij}^{\,\prime\,l}w_{mn}^{\,\prime\,l}=\langle
w_{ij}^{\,\prime},w_{mn}^{\,\prime}\rangle,
\end{eqnarray}
where now
\begin{equation}\label{wmn}
w_{mn}^{\,\prime}=
\left[
\begin{array}{c}
w_{mn}^{\,\prime\,1} \\
w_{mn}^{\,\prime\,2}\\
\vdots\\
w_{mn}^{\,\prime\,K}
\end{array}
\right]=
\left[
\begin{array}{c}
\varphi_1^m\psi_1^n \\
\varphi_2^m\psi_2^n \\
\vdots\\
\varphi_K^m\psi_K^n
\end{array}
\right]=
\left[
\begin{array}{cccc}
\varphi_1^m & & &  \\
&\varphi_2^m  & &  \\
&  & \ddots     &  \\
& & & \varphi_K^m
\end{array}
\right]
\left[
\begin{array}{c}
\psi_1^n \\
\psi_2^n \\
\vdots\\
\psi_K^n
\end{array}
\right]=D_mv_n^{\,\prime},
\end{equation}
\begin{equation}\label{phi}
\phi_l^m:=\bk{\varphi_l}{e_m}, \quad \psi_l^n:=\bk{\psi_l}{f_n}.
\end{equation}

\paragraph*{\bf Reformulation of the separability problem.}
{F}rom (\ref{wmn}) is is thus clear that for a separable state $\rho$ on
$\mathbb{C}^M\otimes\mathbb{C}^N$ which can be decomposed into the sum of $K$
rank-one product operators (\ref{rhos}), there exist $D_m\in\mathbb{M}_{K\times
K}$, $m=1,\ldots,M$, $D_m$ - diagonal, and $v_n^{\,\prime}\in\mathbb{C}^K$,
$n=1,\ldots,N$, such that
\begin{equation}\label{rhoijmn_1}
\rho_{ij,mn}=\langle D_iv_j^{\,\prime},D_mv_n^{\,\prime}\rangle.
\end{equation}

Eq.(\ref{rhoijmn_1}) is also a sufficient condition for separability, ie.\ if
there exist $v_n^{\,\prime}\in\mathbb{C}^K$, $n=1,\ldots,N$ and diagonal
$D_m\in\mathbb{M}_{K\times K}$, $m=1,\ldots,M$ such that (\ref{rhoijmn_1}) is
fulfilled, then $\rho$ can be decomposed into rank-one separable states
(\ref{rhos}) with
\begin{eqnarray}
\ket{\varphi_l}&=&\sum_{m=1}^M\overline{\left(D_m\right)_{ll}}\ket{e_m}, \label{phil} \\
\ket{\psi_l}&=&\sum_{n=1}^N\overline{\left(v_n\right)_l}\ket{f_n}. \label{psil}
\end{eqnarray}
i.e.\ $\rho$ is separable.

Indeed,  define
\begin{equation}
\tilde\rho=\sum_{l=1}^K\kb{\varphi_l\otimes\psi_l}{\varphi_l\otimes\psi_l},
\end{equation}
with $\varphi_l$ and $\psi_l$ defined by (\ref{phil}) and (\ref{psil}). Then
$\tilde\rho$ is  separable and an elementary calculation shows that
$\tilde\rho_{ij,mn}=\rho_{ij,mn}$, and thus $\rho=\tilde\rho$.

Summarizing we can formulate thus the following theorem
\begin{theorem}\label{theorem:sep}
A state $\rho$ is separable if and only if there exists a Gram decomposition of
$\rho$,
$$\rho_{ij,mn}=\langle w_{ij},w_{mn}\rangle,$$
$w_{ij}\in\mathbb{C}^K$ for some $K$, such that
$$ w_{ij}=D_i v_j,$$
with  $N$ vectors $\{v_1,\dots
v_N\}\in\mathbb{C}^K$ and $M$ diagonal matrices $D_1,\dots D_M$
acting as operators on $ \mathbb{C}^K $. \hfill $\Box$
\end{theorem}
Observe that we can assume that all diagonal matrices $D_i$ are nonsingular.
Indeed from (\ref{wmn}) their diagonal elements are equal to the projections of
the vectors $\ket{\phi_{\,l}}$, which constitute (a part of) the decomposition,
onto the basis vectors $\ket{e_m}$. If any number of them vanish we can always
adjust slightly the basis to make them taking non-zero values.

Invoking now Theorem~\ref{theorem:gram1} we obtain a
\begin{corrolary}\label{cor:1}
A state $\rho$ of the full rank $r=MN$ is separable if and only if for some $K\ge MN$,
there exist $K\times r$ matrices $\tilde{V}$ and $V$ of which $\tilde V$ is of
maximal rank and $V^\dagger V=I$, and diagonal $K\times K$ matrices $D_1,\dots
D_M$, such that
$$V^\dagger D_m\tilde{V}=F_m$$
holds on the space spanned by $v_n$, $n=1,\ldots,M$, where $w_{mn}=F_mv_n$ is a
Gram system (\ref{wgen}) for $\rho$ calculated from its spectral decomposition
(\ref{rho}). \hfill $\Box$
\end{corrolary}

Before proceeding let us make a remark. Observe namely that in terms of the
Gram decomposition (\ref{rhoijmn_1}) of a separable $\rho$ the operation of
partial transposition in $\mathcal{H}_B$
\begin{equation}\label{pt1}
\rho=\sum_{i=1}^{k}p_i \rho_i^{A}\otimes \rho_i^B\mapsto\rho^{T_B}
=\sum_{i=1}^{k}p_i \rho_i^{A}\otimes (\rho_i^B)^T,
\end{equation}
ie.\ $\rho_{ij,mn}\mapsto\rho_{in,mj}$,
corresponds to the complex conjugation of the frame $\{v_1,\dots,v_N\}$, ie.
\begin{equation}\label{pt3}
\{v_1,\dots , v_N\}\to \{\overline{v_1},\dots ,\overline{v_N}\},
\end{equation}
whereas a similar operation performed in $\mathcal{H}_A$
\begin{equation}\label{pt1a}
\rho=\sum_{i=1}^{k}p_i \rho_i^{A}\otimes \rho_i^B\mapsto\rho^{T_A}
=\sum_{i=1}^{k}p_i (\rho_i^A)^T\otimes\rho_i^B,
\end{equation}
ie.\ $\rho_{ij,mn}\mapsto\rho_{mj,in}$, consists in
\begin{equation}\label{pt3a}
\{D_1,\dots , D_M\}\to \{\overline{D_1},\dots ,\overline{D_M}\}.
\end{equation}
Indeed:
\begin{eqnarray*}
\rho_{ij,mn}^{T_B}&=&\rho_{in,mj}=\langle D_iv_n,D_mv_j\rangle=
\langle v_n,\overline{D_i}D_mv_j\rangle=\overline{\langle
\overline{D_i}D_mv_j,v_n\rangle}= \langle D_i
\overline{D_m}\overline{v_j},\overline{v_n}\rangle \\
&=&\langle \overline{D_m}D_iv_j,\overline{v_n}\rangle=
\langle D_i\overline{v_j},D_m\overline{v_n}\rangle, \\
\rho_{ij,mn}^{T_A}&=&\rho_{mj,in}=\langle D_mv_j,D_iv_n\rangle=
\langle v_j,\overline{D_m}D_iv_n\rangle=
\langle v_j,D_i\overline{D_m}v_n\rangle=
\langle\overline{D_i} v_j,\overline{D_m}v_n\rangle,
\end{eqnarray*}
where we used the fact that diagonal matrices commute and their Hermitian
conjugation reduces to the complex one.

In Appendix \ref{app:werner}  we discuss in detail the results of this section
applied to an example of two qubit states, the so called Werner states.

\section{Separability and full families of commuting normal matrices}
\paragraph*{\bf $K$-separability and FFCNM.}
In forthcoming publications we will present applications of
Corollary~\ref{cor:1} to characterization of the bipartite entanglement for
arbitrary systems. In the present paper we would like to concentrate on the
separability of systems with $M=2$, but before that we would like to  use our
results from the previous sections to formulate a novel SNC for separability.
Let us assume that the investigated state $\rho$ is $K$-separable, i.e.\ there
exists a decomposition into exactly $K$ rank-one product operators (\ref{rhos})
and $\rho$ can be cast in the from (\ref{rhoijmn_1}). Since we assumed that
$\rho$ is of maximal rank $r=NM$ we have necessarily $K\ge MN$.  From the
previous consideration we now that the Gram vectors calculated with the help of
this decomposition have the form $w_{mn}^{\,\prime}=D_mv_n^{\,\prime}$.

For an arbitrary decomposition of $\rho$ into exactly $K$ states given by
(\ref{rho1}) (where $\ket{\Phi_l}$ need not to be product states), we obtain
another Gram system
$w_{mn}^{\,\prime\prime\,l}=\bk{\Phi_l}{e_m\otimes f_n}\in\mathbb{C}^K$.

The vectors $w_{mn}^{\,\prime}$ and $w_{mn}^{\,\prime\prime}$ as forming two
Gram systems for the same matrix $\rho$ are connected via a unitary
transformation
\begin{equation}\label{M1}
w_{mn}^{\,\prime\prime}=Uw_{mn}^{\,\prime}=UD_mv_n^{\,\prime}.
\end{equation}
Taking the above equality for two pairs of indices $(m,n)$ and $(k,n)$ we
obtain:
\begin{equation}\label{M2}
M_{mk}w_{kn}^{\,\prime\prime}=w_{mn}^{\,\prime\prime},
\end{equation}
where
\begin{equation}\label{M3}
M_{mk}=UD_m(D_k)^{-1}U^\dagger.
\end{equation}
Remember that without loosing generality we can assume nonsingularity of all
matrices $D_n$. Consequently $M_{km}$ are also nonsingular.

\paragraph*{\bf SNC for separability and FFCNM.}
The matrices $M_{nm}$ are normal, $\left[M_{km},M_{km}^\dagger\right]=0$, and
mutually commuting, $\left[M_{km},M_{lm'}\right]=0$. Both observations can be
easily proved using the facts that all matrices $D_m$ are diagonal and $U$ is
unitary. The above reasoning is summarized in the form of the following
\begin{theorem}\label{theorem:norm}
A necessary and sufficient condition for $K$-separability of $\rho$ is the
existence, for an arbitrary decomposition (\ref{rho1}), of a full family of
$M(M-1)/2$ normal, commuting $K\times K$ matrices $M_{km}$ fulfilling
(\ref{M2}) where $w_{mn}^\prime$ are appropriate Gram vectors for the
decomposition (\ref{M2}).
\end{theorem}
Necessity of the condition follows from the above remarks, and to prove the
sufficiency let us assume that (\ref{M2}) is fulfilled for some family of
normal, commuting matrices $M_{km}$. It is a standard fact from the linear
algebra \cite{horn} that all matrices in such a family can be simultaneously
diagonalized by a single unitary transformation,
\begin{equation}\label{M4}
U^\dagger M_{km} U= D_{km}.
\end{equation}
According to the previous remarks we assume that $M_{km}$ and, consequently,
$D_{km}$ are nonsingular. Now from (\ref{M2}) and (\ref{M4})
\begin{equation}\label{M5}
D_{km}U^\dagger w_{mn}^{\,\prime\prime}=U^\dagger M_{km} U U^\dagger
w_{mn}^{\,\prime\prime}=U^\dagger M_{km}w_{mn}^{\,\prime\prime}=U^\dagger
w_{kn}^{\,\prime\prime},
\end{equation}
and defining $v_n:=U^\dagger w_{1n}^{\,\prime\prime}$, $w_{kn}:=U^\dagger
w_{kn}^{\,\prime\prime}$, we obtain
\begin{equation}\label{M6}
w_{kn}=(D_{1k})^{-1}v_n.
\end{equation}
The vectors $w_{kn}$ are Gram vectors for $\rho$ as they are obtained by a
single unitary transformation form the vectors $w_{kn}^{\,\prime\prime}$
constituting some Gram decomposition of $\rho$. Equation (\ref{M5}) reveals
their structure in the form sufficient for the separability of $\rho$ according
to Theorem~\ref{theorem:sep}.

\section{Separability in $2\times N$ systems and normal extensions}

The Theorem 3 simplifies significantly for $2\times N$ systems, because FFCNM
consists of a single matrix, which has to fulfill
\begin{equation}\label{M2a}
\hat M w_{0n}^{\,\prime\prime}=w_{1n}^{\,\prime\prime}.
\end{equation}
In the following we will use ${0,1}$ instead of ${1,2}$ for numbering the
components on the qubit side, which is more in accord with the custom to denote
the basis states by $\ket{0}$ and $\ket{1}$. From here we do not need to assume
the nonsingularity of $\rho$ - see the remarks preceding the formula (\ref{Pm}).

In this section we study the consequences of (\ref{M2a}). On one hand we  use the
present formulation to obtain particularly simple proofs of known separability
criteria. On the other hand, we use known separability criteria to obtain
non-trivial statements concerning existence of normal extensions of matrices,
whose matrix elements are only partially  known.

\paragraph*{\bf Canonical forms and PPT condition.}

Let us consider $\rho$ in the canonical form \cite{2N} (see also Appendix
\ref{app:2xN})
\begin{equation}\label{2Nr-main}
\rho=
\left[
\begin{array}{cc}
 A     & B \\
 B^\dagger & I \\
\end{array}
\right].
\end{equation}
where the positivity of $\rho$ implies $A=BB^\dagger+\Lambda\Lambda^\dagger$,
where $\Lambda$ is some $N\times p$ matrix, with $p\ge
r(\Lambda\Lambda^\dagger)$. Obviously, $p=1$ necessarily, when
$r(\Lambda\Lambda^\dagger)=1$; also one can always take the minimal
$p=r(\Lambda\Lambda^\dagger)=1$.  We represent
$\Lambda\Lambda^\dagger=\sum_{n=1}^p\ket{\Lambda_n}\bra{\Lambda_n}$.

Similar considerations concern the partially transposed matrix, which reads
\begin{equation}\label{pt0-main}
\rho^{T_A}=
\left[
\begin{array}{cc}
 A     & B^\dag \\
 B & I \\
\end{array}
\right].
\end{equation}
We consider here only the nontrivial case of states with the positive partial
transpose (PPT states) - states which are not PPT are not separable. The
positivity of $\rho^{T_A}$ requires now that $A=B^\dagger
B+\tilde\Lambda^\dagger\tilde\Lambda$, where $\tilde\Lambda$ is now a $\tilde
p\times N$ matrix, with $\tilde p\ge r(\tilde\Lambda^{\dag}\tilde\Lambda)$, and
having analogous properties as $p$ introduced above. We represent
$\tilde\Lambda^\dag\tilde\Lambda=\sum_{n=1}^´{\tilde
p}\ket{\tilde\Lambda_n}\bra{\tilde\Lambda_n}$.  The PPT condition can be thus
stated as $A-B^\dagger B\ge 0$. More precisely, it must hold
\begin{equation}
A=BB^\dag+\Lambda\Lambda^\dag=B^\dag B+\tilde\Lambda^\dag\tilde\Lambda,
\label{ppt-mod}
\end{equation}
which implies that given $\Lambda$, $\tilde\Lambda$ are not independent, and
related  by the above constraint.

Let us now discuss several examples to show how the novel entanglement SNC works.

\paragraph*{\bf Rank $N$ matrices.}
The results of Ref.~\cite{2N} indicate that rank $N$ PPT states are
$N$-separable. The matrix $\hat M=B$ then, and $[B,B^{\dag} ]=0$.

\paragraph*{\bf The case $\rho=\rho^{T_A}$.}

{F}rom Ref.~\cite{2N} we gather also that when $\rho=\rho^{T_A}$, then $\rho$ is
$2N$-separable. In this case $B=B^{\dag}$, $\Lambda=\tilde\Lambda^{\dag}$ and
the matrix $M$ can be written as
\begin{equation}\label{M-real-main}
\hat M=
\left[
\begin{array}{cc}
 B     & \Lambda \\
 \Lambda^{\dag} & s \\
\end{array}
\right].
\end{equation}
with $N\times N$ matrix $S$ to be determined. Obviously, taking $s$ Hermitian
provides the desired normal extension of $\hat M$.

\paragraph*{\bf The case of $2\times 2$ and $2\times 3$ systems.}
In the two qubit, or qubit-qutrit case, every separable matrix is
$K$-separable, where $K={\max}(r(\rho),r(\rho^{T_A}))$. In particular for the
full rank $r(\rho)=4$ ($r(\rho)=6$),  $K=4$ \cite{anna} or $K=6$ (as shown in
Appendix B), respectively. We have then
\begin{corrolary}
For $N=2,3$, an arbitrary $N\times N$ matrix $B$, and an arbitrary $p\times N$
matrix $\Lambda$ constrained by (\ref{ppt-mod}), the matrix

\begin{equation}\label{M-2x2-main}
\hat M=
\left[
\begin{array}{cc}
 B     & \Lambda \\
 \tilde\Lambda & s \\
\end{array}
\right].
\end{equation}a
has a normal extension, i.e. there exist a  $p\times p$ matrix $s$, and  a
$N\times p$ matrix $\tilde\Lambda$ constrained by (\ref{ppt-mod})such that
$\hat M$ is normal.  This holds in particular for minimal $p=\min{0,
r(\rho)-N}$.
\end{corrolary}

\paragraph*{\bf Edge PPT entangled states for $N=4$.}
Perhaps the most interesting are applications for PPT entangled states, and in
particular for the so called edge states \cite{2N}, i.e.\ PPT states that
cannot be represented as a mixture of a PPT state and a separable states (no
separable part can be subtracted). Such states are extreme examples of states
to which the range criterion of P. Horodecki \cite{pawel} applies. For $N=4$
such states may have rank 5, or 6 (and similarly their partial transpose). From
the analysis of the Appendix B we infer that if $\rho$ is a separable state of
rank 5 such that its partial transpose has rank 5 (6), then it is 5-separable
(6-separable). In the case $r(\rho)=r(\rho^{T_A})=5$, both $\Lambda$ and
$\tilde\Lambda$ have rank 1; we denote $\Lambda$ by $\ket{\Lambda}$, and
$\tilde\Lambda^{\dag}$ by $\ket{\tilde\Lambda}$. We get then

\begin{corrolary}
A PPT state $\rho$ such that it and its partial transpose have rank 5 is
separable, iff there exist a complex number $s$ such that the matrix
\begin{equation}\label{M-5,5-main}
\hat M=
\left[
\begin{array}{cc}
 B     & \ket{\Lambda} \\
\bra{\tilde\Lambda} & s \\
\end{array}
\right].
\end{equation}
is normal, which assuming that (\ref{ppt-mod}) holds, requires that
$$(B-s)\ket{\tilde\Lambda}=(B^{\dag}-s^*)\ket{\Lambda}.$$
\end{corrolary}
This condition is equivalent to the range criterion. For the particular example
$\rho_{97}$ of the $2\times 4$ state analyzed in the seminal 1997 paper
\cite{pawel}, it is very easy to analyse, as we show in Appendix C.

This analysis may be extended to the rank 5 states, with the partial transpose
of rank 6, for which Eq. (\ref{ppt-mod}) becomes
\begin{equation}
BB^\dag+\ket{\Lambda}\bra{\Lambda^\dag}=B^\dag
B+\ket{\tilde\Lambda^\dag_1}\bra{\tilde\Lambda_1} +
\ket{\tilde\Lambda^\dag_2}\bra{\tilde\Lambda_2}. \label{ppt-mod1}
\end{equation}

We have in this case
\begin{corrolary}
A PPT state $\rho$ of rank 5, such that its partial transpose has  rank 6, is
separable, iff there exist complex numbers $\alpha$, $\beta$, such that
$|\alpha|^2+|\beta|^2=1$, and a $2\times 2$ matrix   $s$ such that the matrix
\begin{equation}\label{M-5,6-main}
\hat M=
\left[
\begin{array}{ccc}
 B     & \alpha\ket{\Lambda} & \beta\ket{\Lambda}\\
\bra{\tilde\Lambda_1} & s_{11} & s_{12} \\
\bra{\tilde\Lambda_2} & s_{21} & s_{22}
\end{array}
\right].
\end{equation}
is normal, which assuming that (\ref{ppt-mod1}) holds, requires that
$$(B-s_{11})\ket{\tilde\Lambda_1}-s_{21}\ket{\tilde\Lambda_2}=
(\alpha B^{\dag}-\alpha s^*_{11}-\beta s^*_{12})\ket{\tilde\Lambda},$$
$$(B-s_{22})\ket{\tilde\Lambda_2}-s_{12}\ket{\tilde\Lambda_1}=
(\beta B^{\dag}-\alpha s^*_{21}-\beta s^*_{22})\ket{\tilde\Lambda}.$$
\end{corrolary}

Before we end this section, we would like to stress that obviously the above
discussion of normal extension of $\hat M$ applies also to $M\times N$ systems,
if we focus on a single relation of the type (\ref{M2}), such as say
\begin{equation}\label{M2-ogol}
\hat M_{10} w_{0n}^{\,\prime\prime}=w_{1n}^{\,\prime\prime}.
\end{equation}
The analysis pertains then to the study of separability on a particular
$2\times N$ subspace of the full Hilbert space. In this sense it is somewhat
similar to the theory of generalized concurrences of Ref.~\cite{kus}.

\section{Summary}

We have presented a new approach to the separability problem by reformulating
it in terms of existence of  separable Gram decompositions of density matrices
in auxiliary space. The existence of such Gram decompositions is equivalent to
the existence of a full family of commuting normal matrices that relate
components of Gram vectors. We have presented many examples and applications of
this method mainly to the $2\times N$ systems. Several known separability
criteria can be, on one hand,  reproduced with this method in a particulary
simple way, and on the other, can be used to derive nontrivial statements about
the existence of FFCNM.

\begin{acknowledgments}
We thank  I. Cirac, F. Hulpke, Ph. Hyllus, J. Korbicz, B. Kraus, and A. Sanpera
for helpful discussions. We acknowledge support of ESF PESC ``QUDEDIS'', EU IP
``SCALA'', Spanish MEC (FIS2005-04627 and Consolider Ingenio 2010 "QOIT"), and
Polish grant PBZ-Min-008/P03/03.
\end{acknowledgments}

\begin{appendix}

\section{Werner matrices for two qubits}
\label{app:werner}

As an example illustrating the results of Section \ref{sec:separ} let us
consider a one-parameter family of states for $N=M=2$
\begin{equation}\label{werner}
\rho=\frac{1}{4} \left[
\begin{array}{cccc}
                    1+p & 0   & 0   & 2p \\
\noalign{\medskip}  0   & 1-p & 0   & 0  \\
\noalign{\medskip}  0   & 0   & 1-p & 0  \\
\noalign{\medskip}  2p  & 0   & 0   & 1+p
\end {array} \right],
\end{equation}
the so called Werner states \cite{werner}. The parameter $p$ takes the values
from the interval $[0,1]$. One finds easily the spectral decomposition
$\rho=\sum_{l=1}^r\kb{\Psi_l}{\Psi_l}$ with
\begin{equation}\label{wernerpsi}
\ket{\Psi_1}=\left[
\begin {array}{c}
0\\
\sqrt{\frac{1-p}{8}}\\
-\sqrt{\frac{1-p}{8}}\\
0
\end{array}
\right],
\ket{\Psi_2}=\left[
\begin {array}{c}
\sqrt{\frac{1-p}{8}}\\
0\\
0\\
-\sqrt{\frac{1-p}{8}}
\end {array}
\right],
\ket{\Psi_3}=\left[
\begin {array}{c}
0\\
\sqrt {\frac{1-p}{8}}\\
\sqrt {\frac{1-p}{8}}\\
0
\end {array}
\right],
\ket{\Psi_4}=\left[
\begin {array}{c}
\sqrt{\frac{1+3\,p}{8}} \\
0 \\
0 \\
\sqrt{\frac{1+3\,p}{8}}
\end {array}
\right],
\end{equation}
and calculates the Gram vectors (\ref{wgen})
\begin{equation}\label{werenergram}
w_{11}=\left[
\begin {array}{c}
0\\
\sqrt{\frac{1-p}{8}}\\
0\\
\sqrt{\frac{1+3\,p}{8}}
\end {array} \right],
w_{12}=\left[
\begin {array}{c}
\sqrt {\frac{1-p}{8}}\\
0\\
\sqrt {\frac{1-p}{8}}\\
0
\end {array}
\right],
w_{21}=\left[
\begin {array}{c}
-\sqrt {\frac{1-p}{8}}\\
0\\
\sqrt {\frac{1-p}{8}}\\
0\end {array}
\right],
w_{22}= \left[
\begin {array}{c}
0\\
\noalign{\medskip}
-\sqrt {\frac{1-p}{8}}\\
\noalign{\medskip}
0\\
\noalign{\medskip}
\sqrt {\frac{1+3\,p}{8}}
\end {array}
\right].
\end{equation}
We chose $v_1=w_{11}$ and $v_2=w_{22}$ which allows to take
\begin{equation}\label{wernerF}
F_1=
\left[\begin {array}{cccc}
1&0&0&0\\
0&1&0&0\\
0&0&1&0\\
0&0&0&1
\end {array}
\right],\quad
F_2=
\left[ \begin {array}{rrcc}
0&-1&0&0\\
-1&0&0&0\\
0&0&0&\sqrt{\frac{1-p}{1+3\,p}}\\
0&0&\sqrt {{\frac {1+3\,p}{1-p}}}&0
\end {array}
\right].
\end{equation}

Only when $p\le 1/3$ the state $\rho$ is separable. For these values of $p$ one
finds an explicit Gram decomposition (\ref{rhoijmn_1}) of $\rho$ with
\begin{equation*}
D_1=\left[\begin{array}{llll}
 1& 0 & 0 & 0 \\
 0& 1& 0& 0\\
 0& 0& 1& 0\\
 0& 0& 0& 1\\
\end{array}
\right],
\end{equation*}
\begin{equation*}
D_2=\left[
\begin{array}{llll}
 \frac{(1-i) \left(\sqrt{1-3 p}+\sqrt{p+1}\right)}{\sqrt{2} \left(\sqrt{1-p}+\sqrt{3 p+1}\right)} & 0 & 0 & 0 \\
 0 & -\frac{(1+i) \left(\sqrt{p+1}-\sqrt{1-3 p}\right)}{\sqrt{2} \left(\sqrt{1-p}-\sqrt{3 p+1}\right)} & 0 & 0 \\
 0 & 0 & -\frac{(1-i) \left(\sqrt{1-3 p}+\sqrt{p+1}\right)}{\sqrt{2} \left(\sqrt{1-p}+\sqrt{3 p+1}\right)} & 0 \\
 0 & 0 & 0 & \frac{(1+i) \left(\sqrt{p+1}-\sqrt{1-3 p}\right)}{\sqrt{2} \left(\sqrt{1-p}-\sqrt{3 p+1}\right)}
\end{array}
 \right],
\end{equation*}
and
\begin{equation*}
v_1^{\,\prime}=\left[\begin{array}{l}
  \frac{ \left(\sqrt{1-p}+\sqrt{3 p+1}\right)}{4
\sqrt{2}}e^{\frac{ \pi i}{2}}\\
 \frac{ \left(\sqrt{1-p}-\sqrt{3 p+1}\right)}{4
\sqrt{2}}e^{\frac{ \pi i}{2}}\\
  \frac{
   \left(\sqrt{1-p}+\sqrt{3 p+1}\right)}{4 \sqrt{2}}e^{-\frac{\pi i}{2}}\\
  \frac{ \left(\sqrt{1-p}-\sqrt{3 p+1}\right)}{4
   \sqrt{2}}e^{-\frac{\pi i}{2}}\\
\end{array}
\right], \quad
v_2^{\,\prime}=\left[\begin{array}{l}
  \frac{ \left(\sqrt{p+1}-\sqrt{1-3 p}\right)}{4 \sqrt{2}}e^{ \frac{  3 \pi i}{4}}\\
  \frac{\left(\sqrt{1-3p}+\sqrt{p+1}\right)}{4 \sqrt{2}}e^{-\frac{ 3 \pi i}{4}}\\
   \frac{\left(\sqrt{p+1}-\sqrt{1-3 p}\right)}{4 \sqrt{2}}e^{\frac{3\pi i}{4}}\\
  \frac{  \left(\sqrt{1-3 p}+\sqrt{p+1}\right)}{4 \sqrt{2}}e^{-\frac{3\pi  i}{4}}
\end{array}
\right].
\end{equation*}

For the particular choice of $F_1,F_2,v_1$, and $v_2$ the matrices $\tilde{V}$
and $V$ (cf.\ Corollary~\ref{cor:1}) are given as
\begin{equation}\label{wernerV}
\tilde{V}=V=\left[ \begin {array}{crcr}
\frac{-\sqrt {1+p}+i\sqrt {1-3\,p}}{\sqrt{8}\sqrt{1-p}}&-\frac{i}{2}&
\frac{\sqrt{1-3\,p}-i\sqrt{1+p}}{\sqrt{8}\sqrt{1-p}}&-\frac{i}{2}\\
\noalign{\medskip}
\frac{-\sqrt {1+p}+i\sqrt {1-3\,p}}{\sqrt{8}\sqrt{1-p}}&-\frac{i}{2}&
\frac{-\sqrt {1-3\,p}+i\sqrt {1+p}}{\sqrt{8}\sqrt{1-p}}&\frac{i}{2}\\
\noalign{\medskip}
\frac{-\sqrt {1+p}+i\sqrt{1-3\,p}}{\sqrt{8}\sqrt {1-p}}&\frac{i}{2}&
\frac{\sqrt {1-3\,p}-i\sqrt {1+p}}{\sqrt{8}\sqrt {1-p}}&\frac{i}{2}\\
\noalign{\medskip}
\frac{-\sqrt {1+p}+i\sqrt{1-3\,p}}{\sqrt{8}\sqrt {1-p}}&\frac{i}{2}&
\frac{-\sqrt {1-3\,p}+i\sqrt {1+p}}{\sqrt{8}\sqrt {1-p}}&-\frac{i}{2}
\end {array} \right],
\end{equation}
for which one easily checks $V^\dagger D_m\tilde{V}=F_m$ on
$span(v_1,v_2)$.

\section{A short guide to $2\times N$ systems}
\label{app:2xN}

{F}rom Theorem \ref{theorem:norm} it is clear that investigations of separability
can be simplified if we know {\it a priori} the order of separability $K$ of
the given state (i.e.\ we know that it is $K$-separable). We do not have any
general tool for determining exactly the order of separability for arbitrary
separable states before finding their actual decomposition into pure products
(and even if we find one, to establish the order of separability we still have
to prove that the found decomposition involves the minimal number of
components). Here we present some exact results concerning orders of
separability in the case of $2\times N$ systems for low values of $N$.

\paragraph*{\bf Canonical forms.}
Let $\rho$ be an arbitrary density matrix of a bipartite $2\times N$ system
\begin{equation}\label{2N}
\rho=
\left[
\begin{array}{cc}
 A     & B \\
 B^\dagger & C \\
\end{array}
\right],
\end{equation}
where $A,B$ and $C$ are $N\times N$ matrices, $A$ and $C$ are hermitian due to
hermiticity of $\rho$. Positive definiteness of $\rho$ implies $A\ge 0$, $C\ge
0$, and $A-BC^{-1}B^\dagger\ge 0$. The matrix $C$ is nonsingular since $\rho$,
by assumption, is of maximal rank. By an invertible transformation
\begin{equation}\label{simp}
\rho\mapsto \left(I\otimes C^{-1/2}\right)\rho\left(I\otimes C^{-1/2}\right)
\end{equation}
we  bring $\rho$ to the {\it canonical form} \cite{2N}
\begin{equation}\label{2Nr}
\rho=
\left[
\begin{array}{cc}
 A     & B \\
 B^\dagger & I \\
\end{array}
\right].
\end{equation}
For simplicity of notation we kept the same symbols $A$ and $B$ to denote the
appropriate blocks of $\rho$ despite the fact that the original blocks
defined in (\ref{2N}) are altered by the transformation (\ref{simp}). Such a
transformation changes, in principle, the trace of $\rho$, but since the
normalization of the trace does not influence separability properties we will
wave this point aside.
The positivity conditions of $\rho$ reduce now to
\begin{equation}\label{ulc}
A=BB^\dagger+\Lambda\Lambda^\dagger,
\end{equation}
where $\Lambda$ is some $p\times N$ matrix.

A necessary criterion of separability is the non-negative definiteness of the
partially transposed matrix, which for the case of a $2\times N$ system (2N) is
defined as
\begin{equation}\label{pt0}
\rho^{T_A}=
\left[
\begin{array}{cc}
 A     & B^\dagger \\
 B & I \\
\end{array}
\right].
\end{equation}

{F}rom now on we will assume thus that both $\rho$ and $\rho^{T_A}$ are
positive-definite (otherwise $\rho$ is not separable). Positive definiteness of
the partial transpose of (\ref{2Nr}) demands
\begin{equation}\label{ulr1}
A=B^\dagger B+\widetilde{\Lambda}^\dagger\widetilde{\Lambda},
\end{equation}
for some $\widetilde{\Lambda}$.

\paragraph*{\bf Decompositions of $\rho$.}
For our purposes it important to consider particular decompositions of $\rho$
for $K=N+p$, and construct the "known" part of the matrix $\hat M$ that
fulfills Eq. (\ref{M2}). From the canonical form and Eq. (\ref{ulc}), with
$\Lambda= (\ket{\Lambda_1},\ldots,\ket{\Lambda_p})$, it is easy to see that
$\rho=\sum_{k=1}^{N+p}\kb{\Phi_k}{\Phi_k}$, with
\begin{equation}
\ket{\Phi_k}=\ket{0}\otimes\ket{k}+ \ket{1}\otimes B\ket{k},
\end{equation}
for $k=1,\ldots, N$, and
\begin{equation}
\ket{\Phi_k}= \ket{1}\otimes\ket{\Lambda_{k-N}},
\end{equation}
for $k=N+1,\ldots, N+p$. From this particular form we read the components of
the vectors $w_{0n}^{\,\prime\prime}, w_{1n}^{\,\prime\prime}$:
\begin{equation}
(w_{0n}^{\,\prime\prime})^k=\delta_{kn},
\end{equation}
for $k=1,\ldots, N$, and zero otherwise. Similarly
\begin{equation}
(w_{1n}^{\,\prime\prime})^k=B_{nk},
\end{equation}
for $k=1,\ldots, N$, and
\begin{equation}
(w_{1n}^{\,\prime\prime})^k=\bk{n}{\Lambda_{k-N}},
\end{equation}
for $k=N+1,\ldots, N+p$. Obviously, for the particularly simple form of
$w_{0n}^{\,\prime\prime}$, Eq.~(\ref{M2}), determines only the first $N$
columns of the matrix $\hat M^T$, which are
\begin{equation}\label{pt0a}
\hat M^{T}=
\left[
\begin{array}{cc}
 B^T     & ? \\
 \Lambda^{\dag} & ? \\
\end{array}
\right],
\end{equation}
where the other entries are at this moment not known. Transposing and using the
PPT constraint (\ref{ppt-mod}), we indeed obtain that
\begin{equation}\label{pt0b}
\hat M=
\left[
\begin{array}{cc}
 B   & \Lambda \\
 \tilde\Lambda & S \\
\end{array}
\right],
\end{equation}
where for given $B$ and $\Lambda$, the matrix $\tilde \Lambda$ is constrained
only by the condition (\ref{ppt-mod}), and $S$ is completely arbitrary.  This
form of $\hat M$ is intensively used by us in the section  on separability in
$2\times N$ systems.

\paragraph*{\bf Edge states.}
For the purpose of this paper we remind the reader the basic concept associated
with the edge states. First, we  remind \cite{LS,2N,karnas} that if
$\ket{e,f}$ (or any vector, in fact) is in the range of $\rho$, then we can
write
$$
\rho= \rho'+\lambda \kb{e,f}{e,f},$$ where $\rho'\ge 0$ provided $\lambda\le
1/\bk{e,f|\rho^{-1}}{e,f}$. When the equality holds, the rank of $\rho'$ is
smaller that the rank of $\rho$ by 1.

{F}rom this observation follows
\begin{corrolary}
A PPT state $\rho$ is an edge state if there exist no product vector
$\ket{e,f}$ in its range, such that $\ket{e^*,f}$ is in the range of
$\rho^{T_A}$.
\end{corrolary}
Subtracting projectors on product vectors as in Refs.\ \cite{LS,2N} allows  to
determine the minimal number of projectors on product states necessary to
decompose a separable state. We remind the reader:

\paragraph*{\bf The case $2\times 2$.}
In this case $K=\max(r(\rho),r(\rho^{T_A}))$. This result stems from
\cite{anna}. In the following we shall use notation $(p,q)$ for the case of
$r(\rho)=p$, $r(\rho^{T_A})=q$. Let us consider the case of full ranks  (4,4).
First we show that we can find the product vector for which  $\lambda=
1/\bk{e,f|\rho^{-1}}{e,f}=1/\bk{e^*,f|(\rho^{T_A})^{-1}}{e^*,f}$, so that
subtracting projector on this vector reduces ranks to (3,3). To this aim we
suppose $\rho=\sum_{k=1}^K\kb{e_k,f_k}{e_k,f_k}$ is separable, and  that for
all $\ket{e,f}$, it holds $\bk{e,f|\rho^{-1}}{e,f} <
\bk{e^*,f|(\rho^{T_A})^{-1}}{e^*,f}$. Inserting into this inequality
$\ket{e_k,f_k}$ and summing over $k$, we get  a contradiction
${Tr}(I)=4<4={Tr}(I)$. In the same manner we prove that the opposite inequality
can not be fulfilled by all product vectors. Thus either all product vectors
fulfill the equality, or there are at least two product vectors for which the
inequality takes opposite signs. But then from the Darboux property and the
fact the the set of all product states is connected, we gather that there exist
a product vector for which the equality holds. In the next step we reduce one
rank to 2; this, however implies that so does the other rank, since from
general theory of Ref.~\cite{2N} it follows that rank $N$ PPT matrix in
$2\times N$ systems is $N$-separable.

\paragraph*{\bf The case $2\times 3$.}
This problem was partially addressed  in the thesis of G. Vidal \cite{vidal}.
The proof here is new. We start with the full ranks and using the same argument
as above we reduce the ranks to (5,5). The argument may be then repeated but
with a certain care. Now we suppose that
$\rho=\sum_{k=1}^K\kb{e_k,f_k}{e_k,f_k}$ is separable, and  that for all
$\ket{e,f}$ in its range, and such that $\ket{e^*,f}$ is in the range of
$\rho^{T_A}$, it holds $\bk{e,f|\rho^{-1}}{e,f} <
\bk{e^*,f|(\rho^{T_A})^{-1}}{e^*,f}$. Again, inserting into this inequality
$\ket{e_k,f_k}$ and summing over $k$, we get  a contradiction ${
Tr}(I_{R(\rho)})=5<5={Tr}(I_{R(\rho^{T_A})})$, where $I_R(\rho)$ denotes
identity on the range. We may again evoke the Darboux property, but to this aim
we need to prove that the set of product vectors on question os connected. Let
$\Psi$ be a vector from the kernel of $\rho$ and $\Phi$ from the kernel of
$\rho^{T_A}$. The product vectors we look for have to fulfill
$\bk{\Psi}{e,f}=0$, $\bk{\Phi}{e,f}=0$. These equations can be regarded as two
linear equations for a three-component vector $\ket{f}$, parametrized by the
vector $\ket{e}=\ket{0}+\alpha\ket{1}$, which we have parametrized by complex
number $\alpha$ in some basis.  Obviosly, $\ket{f}$ is a unique function of
$\alpha$ and by scanning $\alpha$ over the complex plane we can reach any of
these vectors in a continuous way. Darboux theorem says then that there exist a
product vector for which equality holds $\bk{e,f|\rho^{-1}}{e,f} =
\bk{e^*,f|(\rho^{T_A})^{-1}}{e^*,f}$, and we can reduce the ranks to (4,4). The
next step is as above: reduction of one of the ranks to 3, implies the same
reduction for the other. The reason for that is that all  rank 3 states in
$2\times 3$ systems are 3-separable.

\paragraph*{\bf The case $2\times 4$.}
It is also possible to determine what is the minimal number of terms in the
separable decomposition for the states of low ranks. In this paper we consider
two cases: (5,5) and (5,6). In the (5,5) case there are three vectors $\ket{\Psi_i}$ in the kernel of
$\rho$, and another three vectors $\ket{\Phi_i}$ in the kernel of $\rho^{T_A}$.
We look for $\ket{e_k,f_k}$  such that  $\bk{\Psi_i}{e,f}=0$,
$\bk{\Phi_i}{e^*,f}=0$ for all $i=1,2,3$. These can be regarded as six linear
equations for a four-component vector $\ket{f}$. They have solutions provided
three $4\times 4$ determinants (constructed from the first three and one of the
last three equations) vanish. These determinant constitute three polynomials of
3rd order in $\alpha$ and first order in $\alpha^*$. Eliminating $\alpha^*$
from them we obtain that two polynomials of 6th order in $\alpha$ must vanish.
Subtracting them with appropriate coefficients, we conclude that a polynomial
of 5th order in $\alpha$ must vanish, i.e.\ there are at most five product
vectors having the desired properties. This implies that if $\rho$ is
separable, then it is 5-separable.

Similar analysis can be done for the case of the state $\rho$ with the ranks
(5,6). We end up then with one polynomial of 6th order in $\alpha$, i.e.\ we
have at most six solutions, {\it ergo} if $\rho$ is separable, then it is
6-separable. Note, that the states with ranks (5,6) are either separable, or
entangled edge states, or mixtures of ranks (5,5) edge state with a single
projector on a product vector from the range  of $\rho$.

Unfortunately, only upper bounds on the number of product states in an
decomposition of separable states are known for $\rho$'s of higher ranks. In
particular, Caratheodory theorem (for proof see \cite{pawel}) gives a general
bound equal to the square of the dimension of the Hilbert space,  i.e.\ in the
present case $(2\times 4)^2=64$, implying the every separable state is
64-separable.

For the states with ranks (5,7) it can be shown that there exists in the range
of $\rho$ a product vector $(\ket{0}+ \alpha\ket{1})\ket{f}$, such that
$(\ket{0}+ \alpha^*\ket{1})\ket{f}$ is in the range of $\rho^{T_A}$. It is easy
to see that the condition that these product vectors are orthogonal to the
corresponding kernels of $\rho$ and $\rho^{T_A}$, leads to 4 linear equations
for 4 components of $\ket{f}$. The solutions of such equations exits if the
determinant of the corresponding matrix vanishes. This matrix has three rows
linear in $\alpha$ and one row linear in $\alpha^*$, so that the determinant
equations has the form
\begin{equation}
W_3(\alpha)+ \alpha^* V_3(\alpha)=0,
\label{who}
\end{equation}
where $W_3(.)$ and $V_3(.)$ are polynomials of third order. Let us replace
$\alpha\rightarrow rs$, $\alpha^*\rightarrow r/s$ with $r>0$ and $s$ complex,
and treat Eq.~(\ref{who}) as an equation for $s(r)$ (i.e.\ treating $s$ as
parametrically dependent on $r$),
 $$sW_3(rs)+ rV_3(rs)=0.$$
We will show that this equation has at least one root $\alpha= rs$ with
$|s|=1$, i.e. with $\alpha^*= r/s$. To this aim we consider the asymptotic
behavior at $r\rightarrow \infty$. It is easy to show that the above equation
has three roots $s_i=O(1/r)\rightarrow 0$, $i=1,2,3$ and one root
$s_4=O(r)\rightarrow \infty$. Analogously, for $r\rightarrow 0$ it is easy to
show that the equation has three roots $\tilde s_i=O(1/r)\rightarrow \infty$,
$i=1,2,3$ and one root $\tilde s_4=O(r)\rightarrow 0$. All that implies that
when we continuously change $r$ from 0 to $\infty$, one the the three "large"
roots must become "small". From continuity (i.e.\ again from the Darboux
property) we get that for some $r=r_0$, the $|s(r_0)|=1$. Unfortunately, we
cannot say much more about the total number of such roots. Solving
Eq.~(\ref{who}) with respect to $\alpha^*$, complex conjugating the result, and
stacking it back into Eq.~(\ref{who}), we obtain an equation for $\alpha^*$ of
10th order, which indicates that there are not more of than ten roots of
Eq.~(\ref{who}).

\section{Horodecki´s $2\times 4$ edge state}

In this Appendix we show how our method work for the famous state $\rho_{97}$
introduced by P. Horodecki in the seminal paper \cite{pawel}. In our notation
this state has
\begin{equation}\label{B97} B=\left[
                   \begin{array}{cccc}
                   0 & 1 & 0 & 0 \\
                   0 & 0 & 1 & 0 \\
                   0 & 0 & 1 & 0  \\
                   0 & 0 & 0 & 0 \\
                                                 \end{array}
                                               \right]
                                               .
\end{equation}
and
\begin{equation}\label{pawellam}
\ket{\Lambda}=\left[
\begin {array}{c}
\sqrt{(1-b)/2b}\\
0\\
0\\
\sqrt{(1+b)/2b}
\end{array}
\right],
\ket{\tilde\Lambda}=\left[
\begin {array}{c}
\sqrt{(1+b)/2b}\\
0\\
0\\
\sqrt{(1-b)/2b}
\end{array}
\right].
\end{equation}
This state has rank 5 (as its partial transpose) and is an example of an edge
state \cite{2N}. As pointed in Ref.~\cite{pawel} there exist a unitary matrix
$K$, such that $K^2=I$, $KBK=B^{\dag}$, and
$K\ket{\Lambda}=\ket{\tilde\Lambda}$. The condition of existence of the normal
extension from Section VI reads then
$(B-s)\ket{\tilde\Lambda}=K(B-s^*)\ket{\tilde\Lambda}$, i.e.
\begin{equation}\label{paweldec}
\left[
\begin {array}{c}
-s\sqrt{(1+b)/2b}\\
0\\
\sqrt{(1-b)/2b}\\
-s\sqrt{(1-b)/2b}
\end{array}
\right]=
\left[
\begin {array}{c}
-s^*\sqrt{(1-b)/2b}\\
\sqrt{(1-b)/2b}\\
0\\
-s^*\sqrt{(1+b)/2b}
\end{array}
\right],
\end{equation}
which has only the two solutions $s=0$, $b=1$, and the limiting case $b=0$,
with an arbitrary real $s=s^*$. These are exactly the two instances in which the
Horodecki state is separable.
\end{appendix}

\end{document}